\documentclass{epl}

\title{Stabilization of three-dimensional matter-waves solitons in an optical lattice}

\shorttitle{Stabilization of three-dimensional matter-waves...}

\author{Marek Trippenbach\inst{1} \and Micha\l{} Matuszewski\inst{1}\thanks{E-mail:
\email{mmatu@fuw.edu.pl}} \and Boris A. Malomed\inst{2}}

\shortauthor{M. Trippenbach \etal}

\institute{
  \inst{1} Physics Department, Warsaw University - Ho\.{z}a 69, PL-00-681 Warsaw, Poland\\
  \inst{2} Department of Interdisciplinary Sciences, School of
Electrical Engineering, Faculty of Engineering, Tel Aviv University - Tel Aviv 69978,
Israel}




\begin{document}

\maketitle

\begin{abstract}
We propose an experimentally relevant scheme to create stable solitons in a
three-dimensional Bose-Einstein condensate confined by a one-dimensional optical lattice,
using temporal modulation of the scattering length (through ac magnetic field tuned close
to the Feshbach resonance). Another physical interpretation is a possibility to create
stable 3D ``light bullets'' in an optical medium with a longitudinal alternating
self-focusing/defocusing structure, and periodic modulation of the refractive index in a
transverse direction. We develop a variational approximation to identify a stability
region in the parametric space, and verify the existence of stable breathing solitons in
direct simulations. Both methods reveal that stable solitons may be supported if the
average value of the nonlinear coefficient (whose sign corresponds to attraction between
atoms) and the lattice's strength exceed well-defined minimum values. Stable localized
patterns may feature a multi-cell structure.
\end{abstract}

The creation of the Bose--Einstein condensates (BEC) in vapors of alkali metals has
opened an opportunity to investigate nonlinear interactions of atomic matter waves. One
of the most fundamental aspects of these studies is generation of solitons. Dark solitons
had been created in effectively one-dimensional (1D) BEC with repulsive interatomic
collisions~\cite{dark}. Then, bright solitons were observed in 1D BECs with attractive
collisions~\cite{bright,bright2}. An issue of obvious interest is to develop methods for
control of the matter-wave solitons. A promising approach consists in varying the
scattering length (SL)\ of the interatomic collisions by means of an external magnetic
field through the zero-SL point close to the Feshbach resonance~\cite{Inouye}. In
particular, it was shown that an abrupt change of the SL can result in splitting of a
soliton into a set of secondary ones~\cite{Carr}.

The application of ac magnetic field may induce a periodic modulation of the SL, opening
a way to the ``Feshbach-resonance management" (FRM)~\cite{Greece}. Quite a noteworthy
FRM-induced effect is a possibility to create self-trapped oscillating BEC solitons
(breathers) without any external trap. The underlying mechanism is balance between
compression and expansion cycles, corresponding to the time intervals when the SL is,
respectively, negative and positive. In fact, this possibility was first realized in
terms of the transverse light propagation in a bulk layered nonlinear medium, where the
Kerr coefficient alternates between positive and negative values~\cite{Isaac}. The
respective model is formally equivalent to the two-dimensional (2D) Gross-Pitaevskii
equation (GPE), see Eq.~(\ref{NLS}) below, in which the SL\ is periodically modulated in
``time" (analog of the propagation distance in~\cite{Isaac}) not harmonically, but as a
piecewise-constant function. Then, the BEC model based on the GPE with the harmonic
modulation of the SL was directly investigated in Refs.~\cite{Saito,Abdul,VPG}, with a
conclusion that the FRM makes it possible to stabilize, without the use of an external
trap, 2D breathers, but not the 3D ones.
%(contrary to the conclusion of Ref.~\cite{Adhikari})
The absence of the stabilization in the 3D case was also concluded in the optical model
considered in Ref.~\cite{Isaac}, where the 3D soliton would correspond to a
spatiotemporal ``light bullet".

A remarkable feature of these results is that the 2D breather is stable despite a
possibility of the collapse in the 2D GPE with a negative SL. On the other hand, it has
been recently demonstrated that solitons in BECs with a constant negative SL may be
completely stabilized by means of an optical lattice (OL), \textit{i.e.}, a spatially periodic
potential. It is true in \emph{both} 2D and 3D cases~\cite{BBB}. Moreover, it was also
shown that the 2D and 3D BEC solitons can be stabilized by \textit{low-dimensional} OLs,
\textit{i.e}., respectively, by a quasi-1D (Q1D) lattice in the 2D case, and by a quasi-2D (Q2D)
lattice in the 3D space~\cite{Estoril}. However, 3D solitons \emph{cannot} be made stable
in a Q1D OL.

These results suggest a question whether 3D solitons can be stabilized by a combination
of a Q1D lattice and FRM. The issue has practical relevance, as a Q1D OL can be easily
created, illuminating the BEC by a pair of counterpropagating laser beams that form a
periodic interference pattern~\cite{OL}. This problem is different from that considered
in Ref.~\cite{VPG}, where stabilization of a localized 3D structure was provided by
adding a tight 1D parabolic trap to the model, which made it nearly two-dimensional.

In this paper, we demonstrate that the combined OL-FRM stabilization of 3D solitons is
possible indeed. First, we introduce the model and propose an experimental scheme. Then,
we predict stability conditions for the 3D solitons by means of a variational
approximation (VA). Finally, we demonstrate the soliton's stability in direct
simulations. We also demonstrate that these results find another physical realization in
nonlinear optics: a possibility to create a stable 3D ``light bullet" (spatiotemporal
soliton) in a bulk medium which combines an alternating self-focusing/defocusing
structure in the longitudinal direction, and periodic modulation of the refractive index
in one transverse direction.

\section{The model}
\begin{figure}[tbp]
\onefigure[width=9cm]{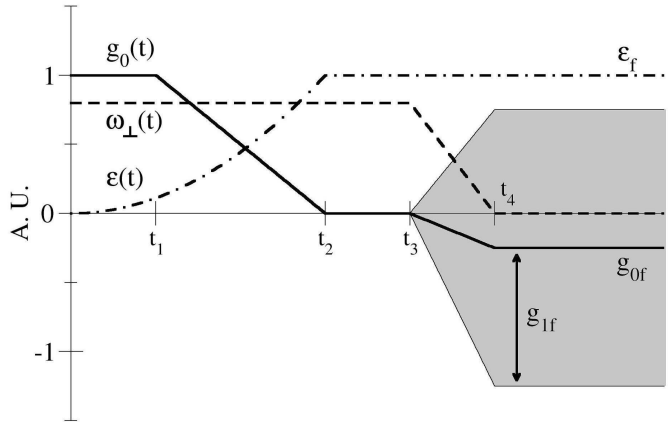} \caption{The time dependence of the nonlinear
coefficient, $g$, transverse-confining frequency, $\protect\omega _{\bot }$, and
optical-lattice strength, $\protect\varepsilon ,$ in the numerical experiment which leads
to the establishment of stable 3D breathing solitons supported by the combination of the
quasi-1D lattice and Feshbach-resonance management (FRM). The shaded area indicates rapid
oscillations of $g$, which account for the FRM.} \label{fig1}
\end{figure}

The GPE for the single-particle wave function $\psi $, including a time-dependent
(FRM-controlled) nonlinear coefficient $g(t)$ and an external potential
$V(\mathbf{r},t)$, in normalized units is
\begin{equation}
i\psi _{t}=\left[ -(1/2)\nabla ^{2}+V(\mathbf{r},t)+g(t)|\psi |^{2}\right] \psi \,.
\label{NLS}
\end{equation} We start by considering a BEC in the ground state of a radial (2D)
parabolic trap with the (time-dependent) frequency $\omega _{\perp }(t)$, supplemented,
in the longitudinal direction, by ``end caps", induced by transverse light sheets. The
configuration is like the one used to create soliton trains in the Li${^{7}}$ condensate
\cite{bright2}. Then, a 1D lattice in the axial direction, whose period is normalized to
be $\pi $, is adiabatically switched on, by increasing its strength $\varepsilon $ from
$0$ at $t=0$ to a final value $\varepsilon_{\ab{f}}$ at $t=t_{2}$, see fig.~\ref{fig1}. Thus,
the full potential is
\begin{equation}
V(\mathbf{r},t)=\varepsilon (t)\left[ 1-\cos (2z)\right] +(1/2)\omega _{\perp
}^{2}(t)\varrho ^{2}+V_{0}(z)\,,  \label{V}
\end{equation}
where $\varrho $ is the radial variable in the plane transverse to $z$, and the axial
end-cap potential $V_{0}(z)$ is approximated by a deep and wide potential box.

At some moment $t=t_{1}<t_{2}$, we begin to linearly decrease an initially positive
nonlinear coefficient $g(t)$. It vanishes at the moment $t=t_{2}$, and remains zero up to
$t=t_{3}$, when we start to gradually switch on the rapid FRM modulation of the nonlinear
coefficient (in fig.~\ref{fig1}, $g(t)$ is denoted as $g_{0}(t)$, up to $t=t_{3}$). At
$t>t_{4}$, $g(t)$ oscillates with a constant amplitude $g_{1\ab{f}}$ around a negative
average value $g_{0\ab{f}}$,
\begin{equation}
g(t)=g_{0\ab{f}}+g_{1\ab{f}}\sin (\Omega t)\,.  \label{g}
\end{equation} Finally, the radial confinement is switched off within the time
interval $t_{3}<t<t_{4}$. After that, a soliton, if any, may only be supported by the
combination of the Q1D lattice and FRM (the axial width of the established soliton is
assumed to be much smaller than the distance between the end caps, hence they do not
affect it).

Numerical experiments following the path outlined in fig.~\ref{fig1} make it possible to
create stable 3D solitons. Before showing the results, we first resort to the VA, in
order to predict conditions on the strength of the OL and size of the negative average
nonlinear coefficient $g_{0\ab{f}}$, which are necessary to support 3D solitons.

Note that the eventual form of Eq.~(\ref{NLS}), \textit{i.e.}, with $g$ and $V$ taken,
respectively, as in Eqs.~(\ref{g}) and (\ref{V}), where $\varepsilon $ is constant and
$\omega _{\perp }=V_{0}(z)=0$, has an alternative physical interpretation in terms of
nonlinear optics. Upon interchanging variables $t$ and coordinates $z$ (chosen to be a
propagation direction) Eq.~(\ref{NLS}) describes pulse propagation in a bulk (3D) medium
composed of alternating self-focusing and self-defocusing layers, similar to that
introduced in Ref.~\cite{Isaac}, with an additional periodic modulation of the refractive
index in the transverse direction $x$. Stable 3D solitons in the present model imply the
existence of stable\emph{\ fully three-dimensional} spatiotemporal solitons (``light
bullets") in this medium. Thus far, only quasi-2D spatiotemporal solitons were created
experimentally in bulk media~\cite{Frank}.

\section{Variational approximation}
The VA was successfully applied to the description of BEC dynamics under diverse
circumstances, see, \textit{e.g.}, Refs.~\cite{VA,Carr,Saito,Abdul,VPG,BBB,Estoril} and a
review~\cite{Progress}. Equation~(\ref{NLS}) is derived from the Lagrangian
\begin{equation}
L=\pi \int \left[ i(\psi _{t}^{\ast }\psi -\psi _{t}^{\ast }\psi )\right. -|\psi
_{\varrho }|^{2}-|\psi _{z}|^{2} -g(t)|\psi |^{4} -\left. 2V|\psi |^{2}\right] \varrho \,
\upd \varrho \, \upd z\,. \label{L}
\end{equation}As it is frequently done, we choose a variational ansatz for the solution
based on the complex Gaussian with an amplitude $A(t)$, radial and axial widths $W(t)\
$and $V(t)$ respectively, $b(t)$ and $\beta (t)$ being the corresponding ``chirps"
\begin{equation}
\psi (\mathbf{r},t)=A\exp\left( -\varrho^{2}\left[1/(2W^{2})+ib\right]
-z^{2}\left[ 1/(2V^{2})+i\beta\right]\right)
\,, \label{ansatz}
\end{equation}
The reduced Lagrangian is obtained by inserting the ansatz~(\ref{ansatz}) into
Eq.~(\ref{L}). The equations obtained by varying the reduced Lagrangian yield the
conservation of the number of atoms, $E\equiv \pi ^{-3/2}\int |\psi |^{2}\upd
\mathbf{r}=A^{2}W^{2}V$, and dynamical equations for the widths,
\begin{eqnarray}
\ddot{W} &=&\frac{1}{W^{3}}-\omega _{\perp }^{2}(t)W+ \frac{Eg(t)}{\sqrt{8}W^{3}V}\,,
\label{variat1} \\
\ddot{V} &=&\frac{1}{V^{3}}-4\varepsilon V\exp \left( -V^{2}\right)
+\frac{Eg(t)}{\sqrt{8}W^{2}V^{2}}\,.  \label{variat2}
\end{eqnarray}

A necessary condition for the existence of a 3D soliton in the present model can be
derived from these equations in a crude approximation, cf. Ref.~\cite{Saito}. To this
end, we assume that $g_{1\ab{f}}$ in Eq.~(\ref{g}) is small, while $\Omega $ is large. It
is also conjectured that the average value $\overline{W}$ of the soliton's radial size,
$W$, is large (see below). Further, in the lowest approximation, the soliton's size in
the axial direction may be assumed constant, $V(t)\approx V_{0}$, as determined by the
relation
\begin{equation}
4\varepsilon V_{0}^{4}\exp \left( -V_{0}^{2}\right) =1\,,  \label{V0}
\end{equation}
that follows from Eq.~(\ref{variat1}) where the last small term ($\sim W^{-2}$) is
dropped. Equation~(\ref{V0}) has real solutions if the OL strength exceeds a minimum
(threshold) value,\begin{equation} \varepsilon \geq \varepsilon
_{\ab{thr}}=e^{2}/16\approx 0.46\,. \label{threshold}
\end{equation}At $\varepsilon >\varepsilon _{\ab{thr}}$, Eq.~(\ref{V0}) has two real
solutions, which implies the existence of two different solitons. It seems very plausible
(cf. Refs.~\cite{BBB,Estoril}) that the narrower one, corresponding to smaller $V_{0}$,
is stable, and the other one is unstable.

Next, replacing $V(t)$ by $V_{0}$ in Eq.~(\ref{variat2}) (we set $\omega _{\perp }=0$, to
consider the possibility of the existence of the 3D soliton without the radial
confinement) and substituting $g(t)$ from Eq.~(\ref{g}), we look for a solution as
$W(t)\approx \overline{W}+W_{1}\sin \left( \Omega t\right) $. For large $\Omega $, the
variable part of the equation yields $W_{1}=-Eg_{1\ab{f}}/\left(
\sqrt{8}\overline{W}^{3}V_{0}\Omega ^{2}\right) $. Then, the consideration of the
constant part of Eq.~(\ref{variat1}), with regard to the first correction generated by
the product of the oscillating terms in $W(t)$ and $g(t)$, produces a
result,\begin{equation} \overline{W}^{4}=\frac{3}{4\sqrt{2}V_{0}}\left(
\frac{Eg_{1\ab{f}}}{\Omega }\right) ^{2}\left( E\left\vert g_{0\ab{f}}\right\vert
-\sqrt{8}V_{0}\right) ^{-1}\,. \label{W0}
\end{equation}An essential corollary of Eq.~(\ref{W0}) is that a necessary condition for
the existence of the 3D soliton is (recall $g_{0\ab{f}}$ is negative)
\begin{equation}
\left\vert g_{0\ab{f}}\right\vert >\left( \left\vert g_{0\ab{f}}\right\vert \right)
_{\min }=\sqrt{8}V_{0}/E\,.  \label{cond}
\end{equation}
Direct simulations presented in the next section show that this condition holds indeed,
albeit approximately. Note that the above approximation, which assumed large
$\overline{W}^{4}$, is valid only when $\left\vert g_{0\ab{f}}\right\vert $ slightly
exceeds the minimum value defined in Eq.~(\ref{cond}).

\section{Numerical results}
\begin{figure}[tbp]
\twofigures[height=4.6cm]{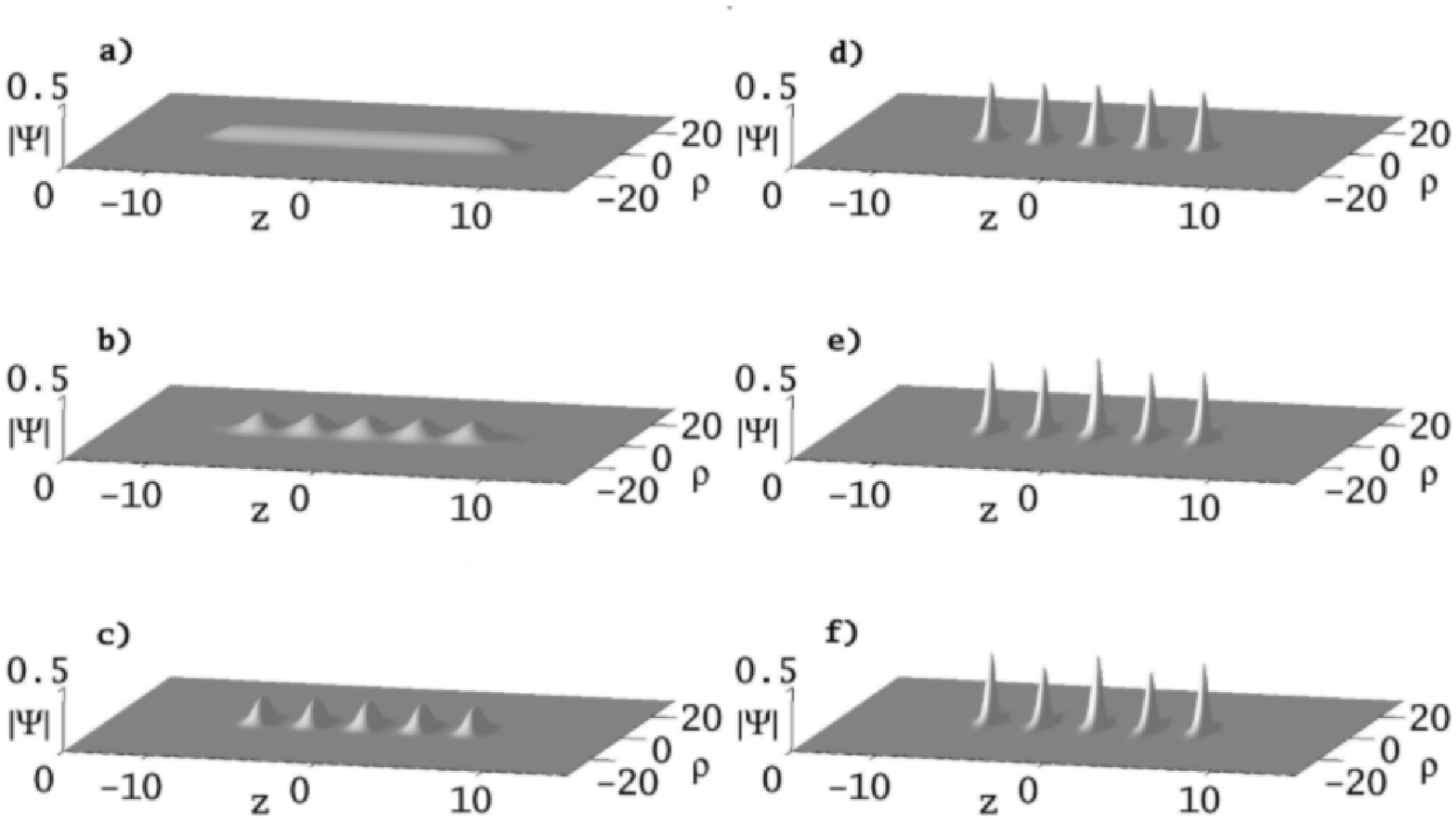}{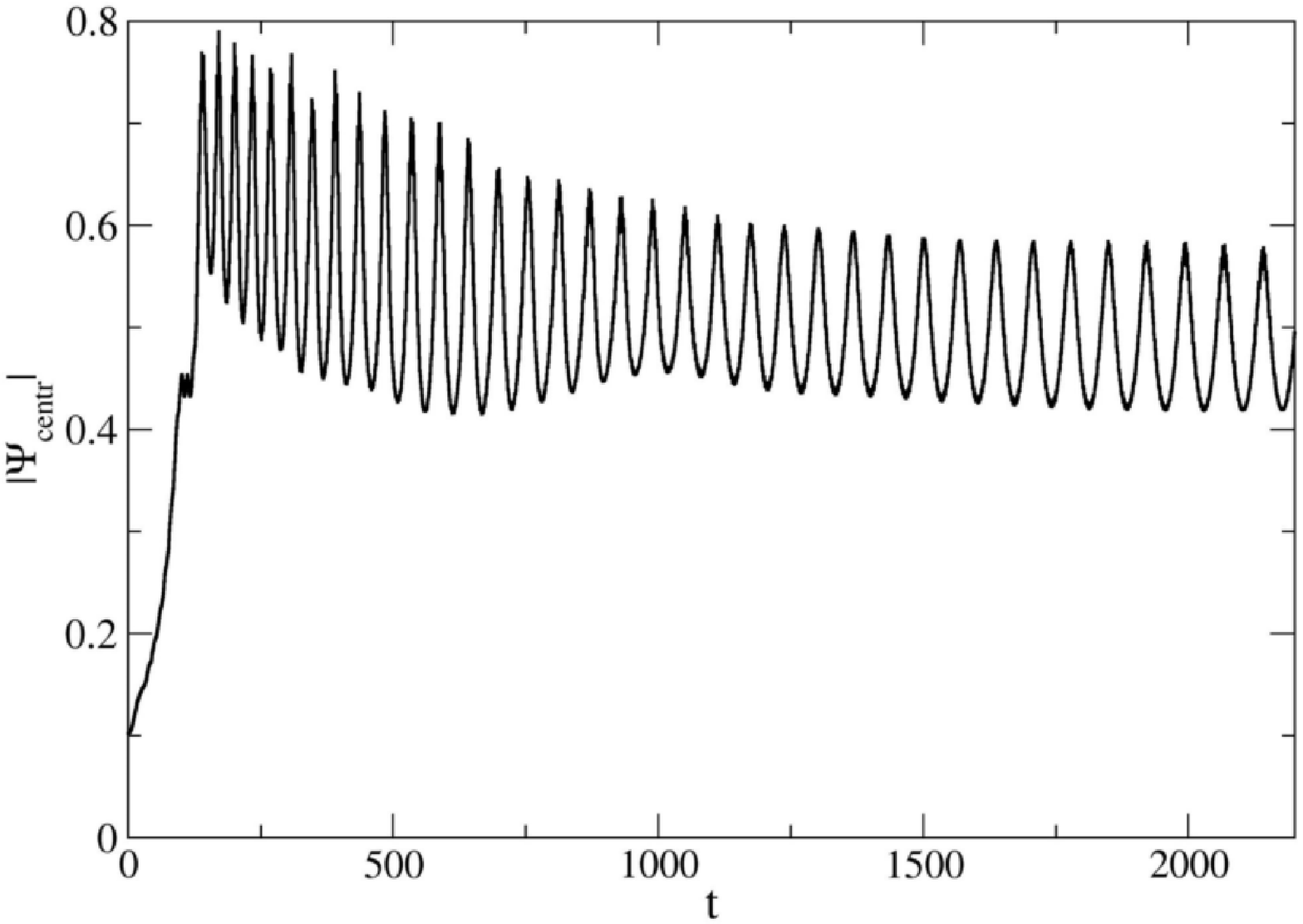}

\caption{Evolution of the absolute value of the wave function in the pattern comprising
five cells of the optical lattice. The normalized parameters are $g_{0\ab{f}}=22.5$,
$g_{1\ab{f}}=4g_{0\ab{f}}$, $\protect\epsilon _{\ab{f}}=25$, $\Omega =40$, $\protect\omega
_{\perp }(0)=0.3$, $t_{1}=30$, $t_{2}=100$, $t_{3}=120$, and $t_{4}=130$. Snapshots are
taken at $t=$ $0$ (a), $20$ (b), $50$ (c) $110$ (d), $200$ (e), $2000$ (f). }
\label{fig2}

\caption{Evolution of the central peak amplitude from fig.~\protect\ref{fig2}.}
\label{fig3}
\end{figure}

We simulated both the full GPE (Eq.~\ref{NLS}) using an axisymmetric code, and the
variational equations~(\ref{variat1} and~\ref{variat2}). In both cases we used
Runge-Kutta methods. In the former case, the simulations followed the path outlined in
fig.~\ref{fig1}. Example of the numerical results, which is quite generic for strong
lattice, is displayed in figs.~\ref{fig2} and~\ref{fig3}. For instance. for $^{87}$Rb
atoms, the OL period is $\lambda =1.5\un{\mu m}$, the FRM and initial radial-confinement
frequencies are, respectively, $\Omega =132\un{kHz}$ and $\omega _{\perp }(0)=990
\un{Hz}$, the lattice depth is $\varepsilon =25$ recoil energies, and the effective
nonlinear coefficient is $na=\pm \,4\cdot 10^{-6}\un{m}$, where $n$ is the number of
atoms per the lattice cell, the total number of atoms being in the range of
$10^{4}-10^{6}$. The respective values of the normalized parameters are given in
fig.~\ref{fig2} caption.

In fig.~\ref{fig2} we show 3D snapshots of the wave-function pattern occupying five
lattice cells, taken at times corresponding to various stages of the numerical
experiment, cf. fig.~\ref{fig1}. Comparison of the last two snapshots, (e) and (f), shows
that condensate develops a robust structure, which remains unchanged over many FRM
cycles.

Figure~\ref{fig3} displays evolution of the central-peak's amplitude in the same pattern.
After an initial transient, the stable structure is established, featuring breathings
without any systematic decay.

%\begin{figure}[tbp]
%\includegraphics[width=8.5cm]{ppotfeshll50.jpg}
%\caption{Evolution of the central peak amplitude from fig.~\protect\ref{fig2}.}
%\label{fig3}
%\end{figure}

Actually, the stable pattern like the one shown in fig.~\ref{fig2} is a set of uncoupled
fundamental solitons, each being trapped in a single lattice cell. This conclusion
follows from additional numerical experiments, in which removal of the atomic population
from any subset of the cells did not, in any tangible way, affect the localized states in
other cells; in particular, the soliton sitting in a single cell is as stable as any
multi-cell pattern. Thus, the conclusion is that the FRM may support a stable 3D soliton
(breather) confined to a \emph{single cell} of the Q1D lattice, and the solitons trapped
in adjacent cells do not interact. In fact, this conclusion complies with the variational
ansatz~(\ref{ansatz}), which ignores any possible spatial oscillations induced by the OL,
hence it indeed implies a soliton essentially confined to a single cell (cf. the VA for
stationary models with the multi-dimensional OLs developed in Refs.~\cite{BBB} and
\cite{Estoril}).

\begin{figure}[tbp]
\onefigure[width=142mm]{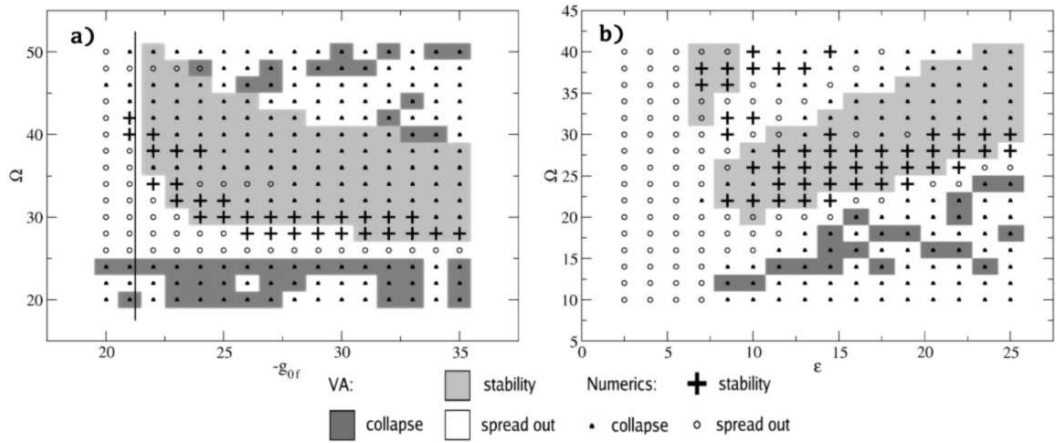}

\caption{Stability regions for the 3D solitons as predicted by the variational
approximation, and found from direct simulations of the Gross-Pitaevskii equation in (a)
the $\left( g_{0\ab{f}},\Omega \right) $ plane, and (b) the $\left( \epsilon,\Omega
\right)$ plane. Other parameters are as in fig.~\protect\ref{fig2}, except that
$g_{0\ab{f}}=30$ in (b). The vertical line in (a) corresponds to the minimum value of
$\left\vert g_{0\ab{f}}\right\vert $ predicted by Eq.~(\protect\ref{cond}).} \label{fig4}
\end{figure}

We have collected results of systematic GPE simulations and compared them with
predictions of the VA based on simulations of Eqs.~(\ref{variat1} and~\ref{variat2}), as
shown in fig.~\ref{fig4}. As is seen in fig.~\ref{fig4}(a), only the bottom part
(corresponding to smaller values of the FRM frequency $\Omega $) of the VA-predicted
stability region in the $\left( g_{0\ab{f}},\Omega \right)$ plane actually
supports stable 3D solitons. The variational estimate~(\ref{cond}) for the minimum size
of the average nonlinear coefficient necessary for the existence of the 3D soliton in the
Q1D lattice is borne out by the simulations, although approximately.

Lastly, the simulations definitely demonstrate the existence of a minimum of the OL
strength $\varepsilon $ which is necessary to support the 3D solitons, as it was
predicted by the VA, see Eq.~(\ref{threshold}). Figure~\ref{fig4}(b) shows regions of
stability in the $\left( \epsilon,\Omega \right)$ plane. From this figure we conclude
that VA predicts stability regions quite well, but the numerical value of the
$\varepsilon _{\ab{thr}}$ is somehow larger than that predicted by the simple
approximation given by Eq.~(\ref{threshold}). For simulations presented in
figs.~\ref{fig2} and~\ref{fig3} we have chosen the value of $\varepsilon$ well above the
threshold.

%\begin{figure}[tbp]
%\includegraphics[width=8.5cm]{ppsdeps.jpg}
%\caption{Stability regions for the 3D solitons in the $\left( \epsilon,\Omega \right)$
%plane, as predicted by the variational approximation, and found from direct simulations
%of the Gross-Pitaevskii equation. Other parameters are as in fig.~\protect\ref{fig2},
%except $g_{0\ab{f}}=30$. We use the same convention as in fig.~\protect\ref{fig4}.}
%\label{fig5}
%\end{figure}

%In fact, this minimum value turns out to be large enough to isolate adjacent cells, which
%explains the above-mentioned fact that that the fundamental solitons trapped in
%neighboring troughs of the OL do not affect each other.

\section{Conclusions} In this work, we have proposed a scheme to stabilize 3D solitons in
BECs where the interatomic interaction is attractive, on average. The scheme is based on
the combination of the Feshbach-resonance management and quasi-1D optical lattice. In
previous works, it has been shown that these two methods applied separately can stabilize
2D solitons, but not 3D ones. The combined method proposed here can be easily implemented
in the experiment, and thus opens a way to the creation of 3D solitons in BECs. Another
physical implication of the results is a stable 3D ``light bullet" in an optical medium
with an alternating longitudinal self-focusing/defocusing structure, and periodic
modulation of the refractive index in a transverse direction.

The possibility to achieve the result was demonstrated within the framework of the
variational approximation (VA) and verified in direct simulations. In particular, the VA
predicts that the OL's strength and average value of the oscillating nonlinear
coefficient must exceed certain minimum values. These predictions are indeed corroborated
by the simulations. However, the full stability region predicted by the VA extends much
farther into the high-frequency region than the actual stability area found in the direct
simulations. The stable patterns found in the simulations may feature a multi-cell
structure, which in the case studied here form a set of non-interacting fundamental
solitons, each trapped in a single cell of the OL, in good qualitative agreement with the
description provided by the VA. In weak lattices one may expect to observe interaction
between individual solitons. This will be a subject of further investigation.

\acknowledgements
M.M. acknowledges support from the KBN grant 2P03 B4325, M.T. was
supported by Polish Ministry of Scientific Research and Information Technology under the
grant PBZ MIN-008/P03/2003. The work of B.A.M. was partially supported by the Israel
Science Foundation through the grant No. 8006/03.

\end{document}